\documentstyle[12pt]{article} 
\def\rs{\rm s} 
\def\rs1{\rm s^{-1}}

\def\rcm{\rm cm} 
\def\rcm2{\rm cm^{-2}}

\def\etal{{\it et al. }} 
 
\def\apjl{{\it Astrophys. J. Lett. }}

\def\nature{{\it Nature }} 
 
\def\aas{{\it Astron. Astrophys. Suppl. Ser.}} 
 
\begin{document} 
\baselineskip=24pt 
  
\centerline{\Large \bf Discovery of the X-Ray Afterglow of the} 
\centerline{\Large \bf Gamma-Ray Burst of February 28 1997} 
\bigskip 
\noindent 
{\it E. Costa$^1$, F. Frontera $^{2,3}$, J. Heise$^4$, 
M. Feroci$^1$, J. in 't Zand$^4$, F. Fiore$^{5,6}$, 
M.N. Cinti$^1$,  D. Dal Fiume$^2$, L. Nicastro$^2$,  M. Orlandini$^2$,   
E. Palazzi$^2$,
M. Rapisarda$^{7,1}$, G. Zavattini$^3$, R. Jager$^4$,  
A. Parmar$^8$, A.Owens$^8$, S.Molendi$^9$, 
G. Cusumano$^{10}$, M.C. Maccarone$^{10}$, S. Giarrusso$^{10}$, 
A. Coletta$^{11}$, L.A. Antonelli$^5$, 
P. Giommi$^5$, J.M. Muller$^{5,4}$,   
L. Piro$^1$, and R.C. Butler$^{12}$}
  
\medskip 
\noindent 
$^1$ Istituto di Astrofisica Spaziale, C.N.R., Via E. Fermi 21, I-00044 
Frascati, Italy 
  
\noindent 
$^2$ Istituto Tecnologie e Studio Radiazioni Extraterrestri, C.N.R., 
Via Gobetti 101, I-40129 Bologna, Italy 
  
\noindent 
$^3$ Dipartimento di Fisica, Universit\`a Ferrara, Via Paradiso 12, 
I-44100 Ferrara, Italy 

\noindent
$^4$ Space Research Organization in the Netherlands,
Sorbonnelaan 2, 3584 CA Utrecht, The Netherlands
 
\noindent
$^5$ Beppo-SAX Scientific Data Center, Via Corcolle 19, I-00131
Roma, Italy
 
\noindent
$^{6}$ Osservatorio Astronomico di Roma, Monteporzio I-00040 Roma, Italy
 
\noindent
$^{7}$ Neutronic Section, Fusion Division, ENEA, Via E. Fermi
45, I-00044 Frascati, Italy
 
\noindent 
$^8$ Astrophysics Division, Space Science Department of ESA, 
ESTEC, P.O.Box 299, 2200 AG Noordwijk, Netherlands 

\noindent
$^9$ Istituto di Fisica Cosmica e Tecnologie Relative, C.N.R., Via
Bassini 15, I-20133 Milano, Italy
 
\noindent 
$^{10}$ Istituto di Fisica Cosmica e Applicazioni Informatica, C.N.R., 
Via U. La Malfa 153, I-90139 Palermo, Italy 
 
\noindent
$^{11}$ Beppo-SAX Science Operation Center, Via Corcolle 19, I-00131
Roma, Italy
 
\noindent 
$^{12}$ Agenzia Spaziale Italiana,  Viale Regina Margherita 202, 
I-00198 Roma, Italy 

\newpage 

{\bf  
The understanding of the nature of gamma-ray bursts  is recognised to be one
of the major challenges of high energy astrophysics. These 
flashes of gamma-rays 
are isotropically distributed in the sky, but  inhomogeneously    
distributed in space, with a deficit of faint 
bursts \cite{fishman95}. It is not yet known if they are produced in our 
galaxy or at cosmological distance.
Only the detection  and identification 
of their counterpart could provide the needed breakthrough
to determine 
the site and the physics of the gamma-ray burst phenomenon. 
Here we report the
discovery in the X-ray band of the first afterglow of a gamma-ray 
burst. It was detected and quickly positioned 
by the Beppo-SAX satellite \cite{boella97a} 
on 1997 February 28 (GRB970228 \cite{costa97a}). 
The X-ray afterglow source was detected \cite{costa97c} with the 
X-ray telescopes aboard the same satellite 
about eight hours after the burst and  faded away in a few days    
with a power law decay function.
The energetic content of the X-ray afterglow results to be 
a significant fraction of gamma-ray burst energetics. 
The Beppo-SAX detection and fast imaging of GRB970228 started a 
multiwavelength campaign that lead to the identification of a fading  
optical  source \cite{Vanparadij97} in a position consistent with 
the X-ray source \cite{fronteraros97}.}

\bigskip                                              

The main reason of our ignorance on the nature of Gamma-Ray Burst (GRB) 
sources is
the unfavourable combination of an unusual phenomenology and instrumental
inadequacy. Gamma-ray telescopes have a poor imaging capability and 
GRBs only last from a fraction of a second to hundreds of seconds. In any
case, after a short time they are no longer detectable in the gamma-ray
band even with large detectors. The burst decay is so fast and the
positioning uncertainty so large that no search for delayed emission
in other wavelengths could so far be successfully 
attempted  \cite{Vrba96}.  

The Italian-Dutch Beppo SAX satellite \cite{piro95} includes many 
experiments in different energy bands and with different
fields of view. In particular, the combined presence of  
an all-sky Gamma-Ray Burst Monitor (GRBM) \cite{costa97b,frontera97}, 
in the 40-700 keV energy range and two Wide Field Cameras
(WFCs) \cite{jager97}, that cover about 5\% of the sky, in the 2-26 keV 
energy range, with a pixel size of 5 arcminutes, 
allows an unprecedented capability of detecting and fast 
positioning GRBs and starting follow-up observations. 

We developed a procedure for fast localization and  
rapid follow-up GRBs observation with the Beppo-SAX Narrow Field
Instruments (NFIs), a cluster of telescopes pointing the same field 
of view and covering the large band  
of 0.1-300~keV \cite{parmar97,boella97b,manzo97,frontera97},   
taking advantage of having them  aboard 
the same satellite and under the same Operation Control Centre. 

On 1997 February 28.123620 UT  the GRBM onboard selection logics 
was triggered by a GRB event.
When the data from the whole orbit were transferred to the ground 
station and forwarded to the Scientific Operation Centre   
quick look analysis of WFCs data at the trigger time showed 
that a counting excess was also present in one WFC. The X-ray 
excess was imaged  showing a point-like source. 
WFC images before and after the event showed that 
the source was transient and simultaneous with the burst.  
Light curves in the gamma-ray and X-ray band are shown in Figure 1.  

The burst position was first determined from a quick look analysis of the
WFC data with an error radius of about 10~arcmin suitable to 
plan a Target of Opportunity (TOO1) pointing of the GRB 
field with Beppo-SAX NFIs.
After few hours through  off-line attitude analysis we obtained for 
the  GRB970228 a refined error box of 3 arcmin radius, centered at
$\alpha\,=\,05^h01^m57^s$, $\delta\,=\,11^\circ46'.4$ (equinox 2000.0).
With these refined position observations in other wavelengths 
were solicited.

The first observation by the Narrow Field Instruments of Beppo-SAX 
started on February 
28.4681, 8 hours only after the GRBM trigger, and ended on 
28.8330. The total exposure time  was 14,344 s in the 
Medium Energy Concentrator Spectrometer and 8,725 s in the Low 
Energy Concentrator Spectrometer. In the refined WFC error box
we found only one source: 1SAX J0501.7+1146 with coordinates 
(equinox 2000.0) $\alpha\,=\,05^h01^m44^s$
and $\delta\,=\,11^\circ46'.7$ and a 90\% confidence error radius
of 50~arcseconds.

Since the pointing of NFIs was based on the first coarse positioning 
of the GRB in two of the three medium 
energy telescopes the source was partially covered by the window
support structure. To exclude spurious 
variability due to pointing drifts, in the analysis we only
use data from the LECS and only one out of three MECS units.
 
The source energy spectrum 
in the 0.1-10~keV band is consistent with a power law of photon index
2.1$\pm$0.3. The hydrogen column density is
$3.5^{+3.3}_{-2.3}\times10^{21}$ cm$^{-2}$ and consistent with
the Galactic absorption along the line of sight $1.6\times10^{21}$ cm$^{-2}$.  
The 2-10 keV average source flux during this
observation was ($2.8\pm 0.4)\times10^{-12}\;\rm\, erg~cm^{-2}~ s^{-1}$,
while the 0.1-2~keV flux was 
$(1.0\pm 0.3)\times10^{-12}\;\rm\, erg~cm^{-2}~s^{-1}$
(note that this corrects the power law photon index, the fluxes and
the hydrogen column density quoted in ref. \cite{Vanparadij97}
).
We also searched for hard X-ray emission (15-100~keV)  
with the Phoswich Detection System  without detecting  
any line or continuum flux. 
The 3$\sigma$ upper limit on the 15-100 keV emission is 
$4.3\times10^{-11}\;\rm\, erg~cm^{-2}~s^{-1}$, which is higher  
than the extrapolation from the low energy power law. 

We performed a second Target of Opportunity (TOO2) observation of the
field with Beppo-SAX NFI, about three days after the
GRB970228 occurrence time (from March 3.7345 to March 4.1174). 
The exposure time was 16,270 s with the MECS and 9,510 s with
the LECS.  A source at a position consistent with that of
1SAX J0501.7+1146 was detected in the MECS. 
Assuming the above spectral shape, the 2-10~keV flux
was $(1.5\pm 0.5)\times10^{-13}\;\rm\,erg~cm^{-2}~s^{-1}$, a
factor about 20 lower than in TOO1.  The source was not detected in the
LECS and the 3 $\sigma$ upper limit in the 0.1-2~keV band was
$4\times10^{-13}\;\rm\, erg~cm^{-2}~s^{-1}$.
In Figure 2 we show the MECS image of the source in the first 
and in the second observation.  
This position is consistent with the GRB error box 
obtained with WFC  and the GRB error annulus resulting from  the 
Interplanetary Network (IPN) based on Beppo-SAX GRBM/Ulysses 
experiments \cite{hurley97c}. 

No source was present in this position in the ROSAT All Sky 
Survey \cite{boller97} with a flux upper limit at 2.5 $\sigma$ 
of $1.9\times10^{-13}\;\rm\, erg~cm^{-2}~s^{-1}$, 
in the range 0.1--2.4 keV, a value compatible with the LECS 
TOO2 but not with TOO1. 

The transient time behaviour and the positional coincidence strongly 
support the association of SAX J0501.7+1146 with GRB970228. 
Using the statistics of X-ray sources derived from the GINGA background 
analysis \cite{warwick} we estimate that the probability to have by chance 
in a field of 3 arcmin radius  a source of intensity 
equal or higher than the one we detected is less than 
8$\times10^{-4}$.  
This probability value is  reduced by at least a 
factor 5 if we take into account the  intersection of the error annulus of 
30~arcsec half-width derived from IPN for GRB970228
\cite{hurley97c,cline97} with the WFC and NFI error boxes. 

While results of a detailed spectral analysis of 1SAXJ0501.7+1146
and GRB970228 will be reported  in another paper (Frontera et al., 
to be submitted), we devote the rest of this letter to the
remarkable time behaviour of the source.  Figure 3 shows the 2-10~keV
flux evolution during the two TOO observations.  
The source flux   shows a significant decrease within the TOO1  
observation . The reduced $\chi^2$ (3 degrees of freedom) for a 
constant  flux is 3.6 corresponding to a probability of 0.13 \%.
We tried to fit data of both observations with a single law. 
An exponential decay function does not fit the data.
The best fit of the TOO1 and TOO2 flux data versus time were obtained with 
a power law function  ($\propto t^{-\alpha}$) (see Fig.3) . 
The best fit  index is given by $\alpha=1.33^{+0.13}_{-0.11}$ 
($\chi^2$ per degree of freedom (dof)~=~0.7  with 4 dof).  

                    
We have also compared the flux and the decay law found 
for 1SAX J0501.7+1146  with the
fluxes measured with GRBM and WFC during the entire gamma-burst 
and during the following minor pulses shown in Fig. 1.  
In Fig. 3 (left top) dashed line shows the 2-10~keV Wide Field Cameras 
flux averaged over 80~s corresponding to the entire burst duration,
while the solid line gives the average  flux of the three minor pulses. 
Both fluxes are consistent with the 
extrapolation of the derived afterglow  decay law.
This strongly suggests that the X-ray emission detected soon after the 
GRB continuously evolves into the X-ray emission of the afterglow. 

This result has a straight implication in terms of the energetics 
of the event.  The GRB fluence  measured by GRBM in 40--700~keV
was $1.1\times10 ^{-5}$ erg cm$^{-2}$. 
The X-ray  fluence measured by WFC in 2--10~keV was about 
$1.2\times10^{-6}$ erg cm$^{-2}$, that is about 11\% of the gamma-ray 
fluence. If we assume that the three last pulses in Fig.1 are part of 
the afterglow  by integrating the power law from 35~s to infinity
we find, in the window 2-10~keV, a fluence which is about 40\% 
of the energy in the gamma burst itself in the band 40--700 keV. 
The X-ray afterglow is not only the low 
energy tail of the GRB phenomenon but it is a significant channel of  
energy dissipation of the event on a completely different 
timescale.

The well established power law decay function of the GRB remnant flux, the 
consistency of its extrapolation with the X-ray flux  at the time of 
the burst, and the energetic content in X-rays are the main results of our
discovery. They will significantly impact on GRB models of 
and constrain their parameters. 
Indeed the fast detection of GRB970228, promptly 
communicated to the scientific community \cite{costa97a,costa97b}, 
triggered both the Beppo-SAX NFI follow-up and 
observations in the radio \cite{galama97a,frail97} and optical bands 
\cite{metzger97,pedersen97,wagner97,tonry97,metzger97b}. 
These observations lead to cogent limits to the radio emission 
and, most, to the detection 
\cite{Vanparadij97,guarnieri97b,margon97,stella97,sahu97,sahu97b} 
of an optical transient, in a position consistent 
with that of 1SAX J0501.7+1146
that faded in a few days. We note, however, that a previous GRB detected
by Beppo-SAX, GRB970111 \cite{970111a}, had a gamma-ray
fluence six times larger than GRB970228 and an undetectable X-ray 
emission 16 hours after the burst. No fading optical source was
detected at a level of magnitude B=23 and R=22.6 \cite{tirado97}. 

The Beppo-SAX measurement, in addition to the discovery of a 
relevant delayed X-ray emission, has 
thus provided the link missing for 25 years between the gamma-ray  
phenomenology and the ultimate location capability of X-ray, optical 
and radio astronomy. 
We expect more future detections of GRBs by Beppo-SAX GRBM/WFC, 
along  with their follow-up observations.  
We hope that the existence of X-ray/optical afterglows and their rapid
detection will contribute to unambiguously identify the GRB sources.

\bigskip

{\bf Acknowledgements}. 
This research is supported by the Italian Space 
Agency (ASI) and Consiglio Nazionale Ricerche of Italy. Beppo-SAX is a joint
program of ASI and the Netherlands Agency for Aerospace Programs (NIVR).
We wish to thank all the people of the Beppo-SAX Scientific Operation 
Centre and the Operation Control Centre for their skilful and enthusiastic 
contribution to the GRB research program.

\newpage 
  
{\bf FIGURE CAPTIONS} 
  
Fig.1. Time profile of GRB970228 in the Gamma-ray (from 
the Gamma Ray Burst Monitor) and X-ray (from the Wide Field Camera) 
bands. The origin is the  trigger time. The   
first pulse is short in  Gamma-rays than in X-rays. Three 
other pulses follow (at around 35, 50 and 70s from the trigger) 
that are much more enhanced in 
the X-ray band. The total burst duration is about 80s.  

Fig.2 Images of the source, 1SAX J0501.7+114, as detected 
with Beppo-SAX  Medium Energy Concentrator Spectrometer (2--10 keV) 
in the error box of GRB90228  during a first  and a second 
Beppo-SAX  Target of Opportunity observation. 
From the ASCA faint sources data \cite{georgantopoulos97} 
the probability that  the source detected during the second pointing  
is coincident by chance with the position of 1SAX J0501.7+1146  
is of the order of $1\times10^{-3}$.
From one pointing to  the other the source is faded by a factor 20 
in three days.

Fig.3. Source flux with time in the 2-10~keV range. Data from the TOO1 
are grouped into 4 points of 8000 s duration each. 
Data from the TOO2 are grouped in one point only due to the lower
statistics. The zero time is taken at the GRBM trigger time. 
Data are fitted by a power law  ($\propto t^{-1.32}$). 
This law is shown as a solid line. The forward extrapolation of the
same law 
is consistent with the flux detected by ASCA \cite{yoshida97}
on March 7.028 of ($8\pm0.3)\times 10^{-14}$ erg cm$^{-2}$ s$^{-1}$ 
(averaged value for SIS and GIS detectors), in same energy range. 

The same law extrapolated backward at the time of the GRB
(described by arrows in the left top) is well matched with the average flux of 
$2.3\times10^{-8}\;\rm\, erg\; cm^{-2} s^{-1}$
detected by WFC in the three minor pulses of Fig.1 from 35 to 70~s. 

Also shown is the 3$\sigma$ upper limit of the source flux 
obtained with WFC 5000 s after the burst for an 
exposure time to the source of 1000 s. 
  
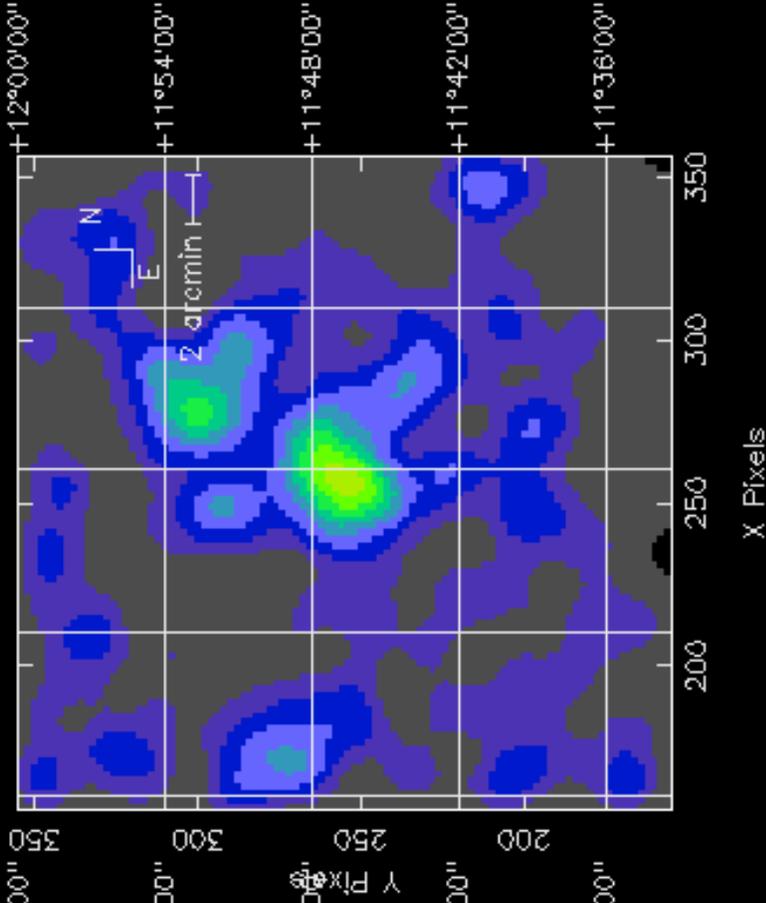
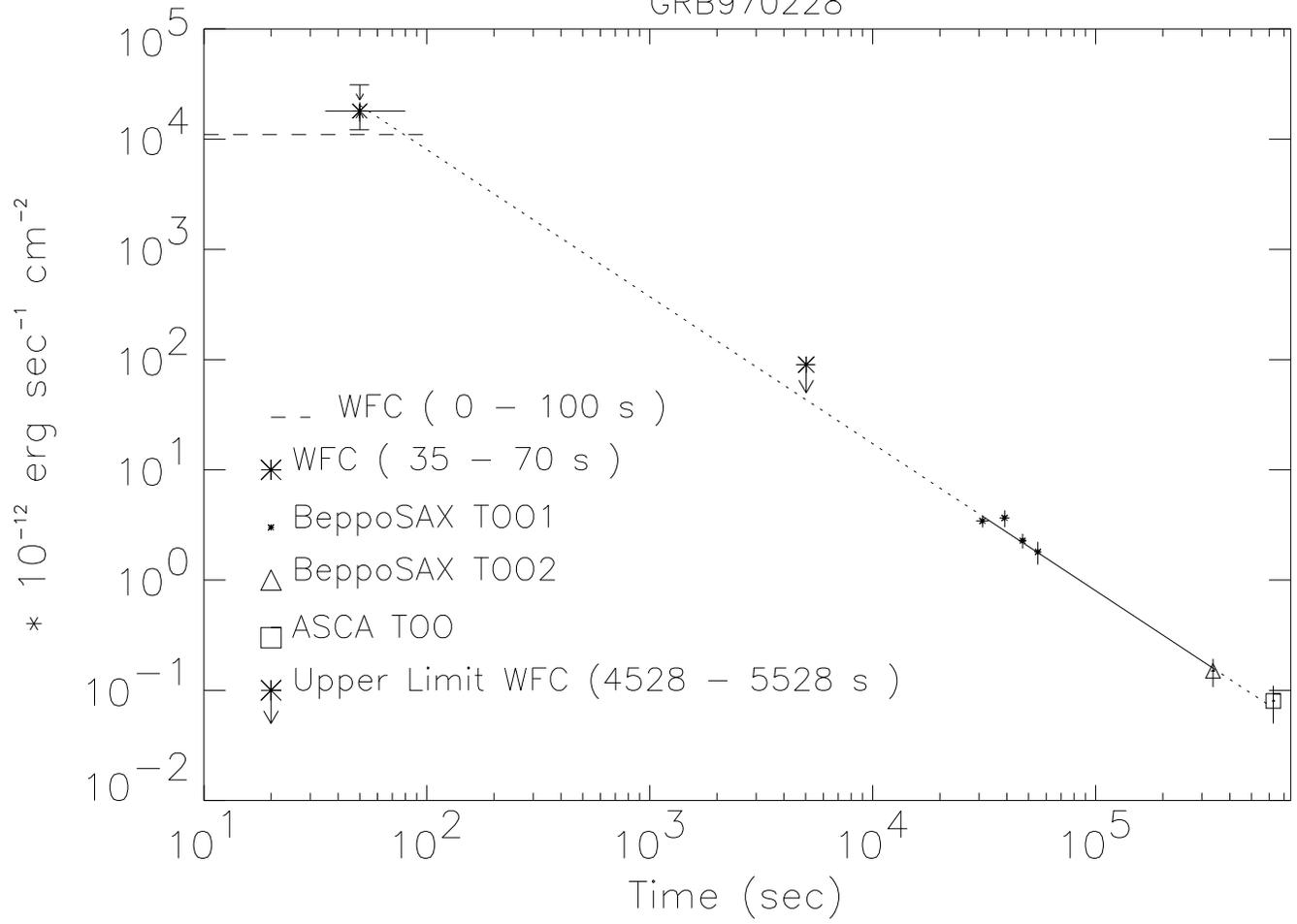
\end{document}